\documentclass[11pt, a4paper]{article}
\usepackage[utf8]{inputenc}
\usepackage{amsmath}
\usepackage{graphicx}
\usepackage{amsfonts}
\usepackage{amssymb}
\usepackage{epsfig}
\usepackage{color}
\usepackage{psfrag}
\usepackage{epstopdf}
\usepackage{authblk}
\usepackage{floatrow}

\newfloatcommand{capbtabbox}{figure}[][\FBwidth]
\newcommand{\EQ}{\begin{equation}}
\newcommand{\EN}{\end{equation}}
\newcommand{\be}{\begin{equation}}
\newcommand{\ee}{\end{equation}}
\newcommand{\bea}{\begin{eqnarray}}
\newcommand{\eea}{\end{eqnarray}}

\title{\bf{On the theory of quantum quenches in near-critical systems}}
\author[1,2]{Gesualdo Delfino\thanks{delfino@sissa.it}}
\author[3]{Jacopo Viti\thanks{jacopo.viti@ect.ufrn.br}}
\affil[1]{\textit{SISSA -- Via Bonomea 265, 34136 Trieste, Italy.}}
\affil[2]{\textit{INFN sezione di Trieste, Italy.}}
\affil[3]{\textit{ECT \& Instituto Internacional
de Fisica, UFRN, Lagoa Nova 59078-970 Natal, Brazil.}}

\date{}
\begin{document}
\maketitle
\begin{abstract}
 \noindent
The theory of quantum quenches in near-critical one-dimensional systems formulated in \cite{quench} yields analytic predictions for the dynamics, unveils a qualitative difference between non-interacting and interacting systems, with undamped oscillations of one-point functions occurring only in the latter case, and explains the presence and role of different time scales. Here we examine additional aspects, determining in particular the relaxation value of one-point functions for small quenches. For a class of quenches we relate this value to the scaling dimensions of the operators. We argue that the $E_8$ spectrum of the Ising chain can be more accessible through a quench than at equilibrium, while for a quench of the plane anisotropy in the XYZ chain we obtain that the one-point function of the quench operator switches from damped to undamped oscillations at $\Delta=1/2$. 
\end{abstract}

\section{Introduction}
The out of equilibrium dynamics of isolated quantum many body systems is in the last years the object of extensive experimental, numerical and theoretical investigations (see e.g. \cite{PSSV,GE,dAKPR}). In particular, the one-dimensional case has attracted special attention due to experimental observations \cite{KWW} and theoretical proposals \cite{RDYO} suggesting that quantum integrability, a well known property of many one-dimensional systems at equilibrium, may affect in a specific way the non-equilibrium dynamics (see \cite{CEM}). On the other hand, the nature of the problem and the many different instances in which it can be considered make difficult to go beyond a case by case analysis and to address with the desired degree of generality questions such as the differences between interacting and non-interacting systems, the role of integrability and the extent of exact solvability, the dependence on initial conditions.

In this context, it is particularly interesting that the theory of quantum quenches in one-dimensional near-critical systems has been formulated in \cite{quench}. A quantum quench is arguably the simplest way to drive a system out of equilibrium. The system is left in an eigenstate (the ground state unless otherwise specified) of its Hamiltonian $H_0$ until the time $t=0$, when the sudden change of a coupling leads to a new Hamiltonian $H$ according to which the system evolves unitarily for positive times. It has been shown in \cite{quench} how, for homogeneous systems on the infinite line near a quantum critical point with emergent relativistic invariance, this problem admits a field theoretical formulation able to answer main questions and to produce analytic predictions. A key point is that the field theoretical description can be performed directly and generally in terms of the particle excitations which are the fundamental dynamical degrees of freedom near criticality. Casting the quenching process into this framework then gives access to the role played in the non-equilibrium dynamics by properties such as the interaction among the particles and the symmetries characteristic of the different universality classes of quantum critical behavior.
A first outcome is that the only exactly solvable cases within the class in exam (which admits the scaling limit both before and after the quench) are those which involve no interaction among the particles, a conclusion which contrasts with the frequency of exact solvability at equilibrium in one dimension; for quantum quenches near criticality, exact solvability means that a connection between the relativistic particles of the pre-quench theory and those of the post-quench theory can be determined exactly. In presence of interaction the theory of \cite{quench} yields results for the post-quench dynamics order by order in the quench parameter $\lambda$ ($\lambda=0$ corresponds to no quench). 

%The expansion in $\lambda$ can be carried through analytically around the different integrable directions in the parameter space of the pre-quench theory and allows to study the dynamics in regions where the theory is strongly interacting.
 
It is a consequence of the theory that the problem possesses two time scales. The first is the pre-quench time scale, inversely proportional to the finite mass gap $m$ of the pre-quench theory. The second time scale $t_\lambda$ is set in a specific way (see (\ref{tlambda}) below) by the quench parameter and goes to infinity as $\lambda$ goes to zero (limit of small quench). When studying perturbatively in $\lambda$ the post-quench dynamics in the long time limit, long time means $1/m\ll t\ll t_\lambda$. We refer to this interval as the {\it intermediate time window}; it is worth stressing that it can be made arbitrarily large taking $\lambda$ sufficiently small. The theory yields already at first order in $\lambda$ a qualitative difference between the interacting and non-interacting cases, showing that undamped oscillations of one-point functions of Lorentz invariant local operators in the intermediate time window can occur in the former case but not in the latter; the oscillation frequencies coincide with the masses of the particles. Taking $\lambda$ small enough, the time scale up to which the theory ensures that the oscillations stay undamped can be made larger 
 than any experimentally or numerically attainable time scale. On the other hand, since there is no exact solution (undamped oscillations can occur only in the interacting case), it is not possible to follow analytically the dynamics in the infinite time limit for fixed $\lambda$. The Ising quantum spin chain with a quench of the longitudinal field was pointed out in \cite{quench} as the basic case where to look for undamped oscillations, and these have indeed been observed in recent numerical studies \cite{KCTC,RMCKT}. The theory also shows that damping of oscillations requires internal symmetries and determines the damping power.

The perturbation theory in $\lambda$ can be formally carried through for a generic near-critical quench. As a consequence, the predictions for properties such as the behavior of one-point functions in the intermediate time window, which turn out to be ruled by internal symmetries, hold in general. On the other hand, if the pre-quench theory is integrable, i.e. possesses an infinite number of conserved currents, the matrix elements (form factors) entering the expansion in $\lambda$ are exactly known and the theory yields analytic results for the full time evolution up to $t_\lambda$. Starting from the different integrable directions in the parameter space of the pre-quench theory, the post-quench dynamics can be followed analytically in time in regions where the theory is strongly interacting. The notion of integrability, that we only use for the equilibrium case, is well defined for the pre-quench theory, which enters the formalism as an equilibrium theory on which the quench acts as a perturbation. It also makes sense to distinguish whether the equilibrium theory corresponding to the post-quench values of the couplings is integrable or not. The formalism, however, applies to both cases, without qualitative differences depending specifically on this feature.

In this paper we investigate additional aspects and implications of the theory. In the first place we determine the time dependence of one-point functions up to $t_\lambda$ for small quenches, including the relaxation value for $t\to t_\lambda\to\infty$. In particular, for small quenches of the mass scale we relate the relaxation value to the scaling dimension of the operator. We then illustrate through a number of interesting examples how our general formulae account for a variety of different situations depending on the pre-quench theory, the quench parameter and the observable. Six qualitatively different quenches are discussed for the Ising chain, comparing the predictions of the theory with available analytical (for free cases) and numerical results, and pointing out interesting implications. In particular, we argue that the correspondence between oscillation frequencies and particle masses, together with the isolation of single-particle states by the long time dynamics, open the way to a more complete observation within the quench setting of the ``$E_8$'' mass spectrum predicted
\cite{Taniguchi} for the Ising chain at critical transverse field and non-zero longitudinal field; this spectrum is very partially observed at equilibrium \cite{Coldea}. Changing system but relying on the same general theory, we find that in a small quench of the plane anisotropy within the XYZ quantum spin chain the one-point function of the quench operator switches from damped to undamped oscillations at the value $\Delta=1/2$ of the second anisotropy parameter, a circumstance that we relate to the properties of the mass spectrum of the sine-Gordon quantum field theory.  We finally approach the issue of the dependence on the initial state, performing an analysis of the case in which the quench is performed from the first excited state of $H_0$ rather than from its ground state.

The paper is organized as follows. In the next section we recall the derivation of \cite{quench} and examine the relaxation value of one-point functions. Quenches in the Ising and XYZ spin chains are then analyzed in section~3, before considering quenches from an excited state in section~4. Section~5 contains few final remarks.

\section{Field theory of quantum quenches near criticality}
Following \cite{quench}, we consider a homogeneous one-dimensional system close to a quantum critical point. Before the quench its scaling limit is described by a quantum field theory with action
\EQ
{\cal A}_0={\cal A}_\textrm{QCP}-g\int_{-\infty}^\infty dt\int_{-\infty}^\infty dx\,\varphi(x,t)\,,
\label{A0}
\EN
where ${\cal A}_\textrm{QCP}$ is the conformally invariant critical point action, and $\varphi$ is the operator which drives the system away from criticality. The field theory in presence of the quench is then specified by the action
\EQ
{\cal A}={\cal A}_0-\lambda\int_0^\infty dt\int_{-\infty}^{\infty} dx\,\Psi(x,t)\,,
\label{action}
\EN
where $\lambda$ and $\Psi$ are the quench parameter and the quench operator, respectively. Both $\varphi$ and $\Psi$ are relevant (or marginally relevant) operators in the renormalization group sense, i.e. their scaling dimensions $X_\varphi$ and $X_\Psi$ at the quantum critical point are smaller than (or, for marginal relevance, equal to) 2. 

Concerning the possibility that the quenched theory (\ref{action}) is exactly solvable, it requires as a precondition that the theory (\ref{A0}), which is an equilibrium theory, is integrable. This means \cite{ZZ} that the scattering theory of (\ref{A0}) is completely elastic (the final state is kinematically identical to the initial one) and factorized. Factorization is related to the fact that the presence of non-trivial conserved currents allows to displace trajectories of particles with different momenta by different amounts in space-time, resulting into the possibility to write the scattering amplitudes of $n$-particle processes as the product of two-body amplitudes. Factorization, i.e. the reduction of the scattering problem to the determination of a finite number of elementary amplitudes, is essential in the process of exact solution and has to hold also for the quenched theory (\ref{action}) if this has to be exactly solvable in the only sense presently known for quantum field theories in $1+1$ dimensions. For the theory (\ref{action}), on the other hand, it is clear that the breaking of time translation invariance allows for creation and destruction of particles at $t=0$ (at least). One could hope for cases in which this happens in peculiar ways compatible with factorization and solvability. However, it is not difficult to see \cite{quench} (see also \cite{Schuricht}) that for the quench process, which allows for particles transmitted from negative to positive times together with particles created and destroyed at $t=0$, factorization is lost unless the particles are free both before and after the quench. Hence this argument leads to the conclusion that, near criticality, quantum quenches are not exactly solvable in presence of interaction\footnote{The very specific way to reconcile factorization and interaction in \cite{Cubero} involves a small parameter $1/N$.}. The argument leaves room for mass quenches in free theories (i.e. ${\cal A}_0$ is a free theory and $\lambda\Psi$ is proportional to its mass term), and these are the cases, bosonic or fermionic, for which the mode dynamics is exactly solved by Bogoliubov transformations \cite{CC,RSMS}. The argument does not exclude solvability for interacting cases which do not admit a scaling limit. For example, the case discussed in \cite{Wouters,Pozsgay} corresponds to a pre-quench Hamiltonian which is far from criticality. More generally, the approach (see e.g. \cite{Wouters,Pozsgay,FCEC,IdNWCEP}) which assumes the existence of a steady state at infinite time and aims at characterizing such a state through a generalized Gibbs ensemble \cite{RDYO} does not address the problem of evolution at finite times we are concerned with. Relaxation dynamics has been investigated analytically in \cite{dNPC} for the Lieb-Liniger model, which however is non-relativistic and does not fall in the class we consider in this paper.

In the near-critical interacting case the theory has to proceed perturbatively, and the perturbation theory in the quench parameter $\lambda$ has been formulated in \cite{quench}. The eigenstates of the pre-quench Hamiltonian $H_0$ are the asymptotic states $|p_1,\ldots,p_n\rangle_{in\,(out)}$ of the equilibrium theory (\ref{A0}), with eigenvalues given by the sum of the particle energies $E_{p_i}=\sqrt{m^2+p_i^2}$. For the sake of notational simplicity we are referring to the case of a single particle species, and $m$ is the particle mass; generalizations will be considered when relevant. In presence of the quench, an initial state $|p_1,\ldots,p_n\rangle_{in}$ at $t=-\infty$ evolves at $t=+\infty$ into a final state that we expand on the basis of outgoing states of ${\cal A}_0$. The coefficients of this expansion are
\EQ
{}_{out}\langle q_1,\ldots,q_m|S_\lambda|p_1,\ldots,p_n\rangle_{in}\,,
\label{coeff}
\EN
with 
\EQ
S_\lambda=T\,\exp\left(-i\lambda\int_0^\infty dt\int_{-\infty}^\infty dx\,\Psi(x,t)\right)
\label{Slambda}
\EN
($T$ denotes chronological ordering); in absence of quench $S_{\lambda=0}=I$ and (\ref{coeff}) reduces to the scattering matrix of the theory (\ref{A0}). 

The simplest and most relevant case is that in which the system is before the quench in the ground state of $H_0$; this coincides with the state without particles, i.e. the vacuum $|0\rangle$. To first order\footnote{The fact that the operator $\Psi$ is relevant or marginally relevant in the renormalization group sense is expected to make the theory (\ref{action}) (super)renormalizable and the small $\lambda$ expansion reliable.} in $\lambda$, it evolves after the quench into the state
\EQ
|\psi_0\rangle=S_\lambda|0\rangle\simeq |0\rangle+\lambda\sum_{n=1}^\infty\frac{2\pi}{n!}\int_{-\infty}^{\infty}\prod_{i=1}^n\frac{dp_i}{2\pi E_{p_i}}\,\delta(\sum_{i=1}^np_i)\,\frac{[F_n^\Psi(p_1,\ldots,p_n)]^*}{\sum_{i=1}^nE_{p_i}}\,|p_1,\ldots,p_n\rangle\,,
\label{psi0}
\EN
where we used $\Psi(x,t)=e^{iP_0x+iH_0t}\Psi(0,0)e^{-iP_0x-iH_0t}$, and the matrix elements
\EQ
F_n^\Psi(p_1,\ldots,p_n)=\langle 0|\Psi(0,0)|p_1,\ldots,p_n\rangle
\label{ff}
\EN
are the form factors of the theory (\ref{A0}). An infinitesimal imaginary part is given to the energy to make the time integral in (\ref{Slambda}) convergent, and leads to the factor $\sum_{i}E_{p_i}$ in the denominator of (\ref{psi0}); some additional details about the derivation are given in \cite{quench}. Here we stress that for free theories (${\cal A}_0$ free and $\Psi$ quadratic, so that $F_n^\Psi\propto\delta_{n,2}$) the sum in (\ref{psi0}) reduces to the contribution of $|p,-p\rangle$; on the other hand, in presence of interaction, (\ref{psi0}) shows that the form consisting of particle pairs with opposite momenta, often considered in the non-equilibrium context, does not occur for quenches near criticality.

We denote by $\delta\langle\Phi(t)\rangle$ the variation of the one-point function of a hermitian operator $\Phi$ with respect to the pre-quench value. At first order in $\lambda$ it reads
\bea
\delta\langle\Phi(t)\rangle &\simeq &{\langle\psi_0|\Phi(x,t)|\psi_0\rangle-\langle 0|\Phi(0,0)|0\rangle}+C_\Phi\nonumber\\
&=& \lambda\sum_{n=1}^\infty\frac{2\pi}{n!}\int_{-\infty}^{\infty}\prod_{j=1}^n\frac{dp_j}{2\pi E_{p_j}}\,\frac{\delta(\sum_{j=1}^np_j)}{\sum_{j=1}^nE_{p_j}}\,\nonumber\\
&\times& 2\mbox{Re}\{[F_n^\Psi(p_1,\ldots,p_n)]^*F_n^\Phi(p_1,\ldots,p_n)\,e^{-i\sum_{j=1}^nE_{p_j}t}\}+C_{\Phi}\,,
\label{1point}
\eea
where we took into account that $\langle 0|0\rangle=1$ and that $\langle\psi_0|\psi_0\rangle=1+{O}(\lambda^2)$. Equation (\ref{1point}) was written in \cite{quench} without the constant
\EQ
C_{\Phi}\equiv -\lambda\sum_{n=1}^\infty\frac{2\pi}{n!}\int_{-\infty}^{\infty}\prod_{j=1}^n\frac{dp_j}{2\pi E_{p_j}}\,\frac{\delta(\sum_{j=1}^np_j)}{\sum_{j=1}^nE_{p_j}}\,2\mbox{Re}\{[F_n^\Psi(p_1,\ldots,p_n)]^*F_n^\Phi(p_1,\ldots,p_n)\}\,,
\label{C}
\EN
which we now introduce requiring the continuity of the one-point function at $t=0$, i.e. $\delta\langle\Phi(0)\rangle=0$, which is not automatic in the scattering framework. For large times the exponential in (\ref{1point}) rapidly oscillates and the integrals are dominated by the contribution of small momenta. For particles with fermionic statistics, which is generic in interacting one-dimensional systems, and for $\Phi$ scalar, the factor $[F_n^\Psi]^*F_n^\Phi$ in (\ref{1point}), evaluated for momenta all tending to zero, will be proportional to $\prod_{1\leq i<k\leq n}(p_i-p_k)^2$, so that the $n$-particle integral will have the large time behavior
\EQ
\frac{\lambda}{t^{(n^2-1)/2}}\,\mbox{Re}(A_n^{\Phi,\Psi}\,e^{-inmt})\,,
\label{leading}
\EN
where $A_n^{\Phi,\Psi}$ are constants. As a consequence we have
\EQ
\delta\langle\Phi(t)\rangle\simeq\frac{\lambda}{t^{(n_0^2-1)/2}}\,\mbox{Re}(A_{n_0}^{\Phi,\Psi}\,e^{-in_0mt})+C_{\Phi}\,,\hspace{.6cm}1/m\ll t\ll t_\lambda\,,
\label{intermediate}
\EN
where $n_0$ is the smallest $n$ for which $[F_n^\Psi]^*F_n^\Phi\neq 0$. In writing (\ref{intermediate}) we also took into account that the perturbative result holds up to a timescale 
\EQ
t_\lambda\sim 1/\lambda^{1/(2-X_\Psi)}\,,
\label{tlambda}
\EN
in such a way that $t/t_\lambda$ remains small at small $\lambda$; we already remarked that $t_\lambda\to\infty$ as $\lambda\to 0$. 

If $n_0=1$ (\ref{intermediate}) exhibits oscillations which are undamped within the intermediate time window. Since, for the scalar operators we consider, one-particle form factors do not depend on momenta, the calculation of the $n=1$ term of (\ref{1point}) is explicitly performed for all times, so that in this case the r.h.s. of (\ref{intermediate}) reads
\EQ
\lambda\sum_a\frac{2}{m_a^2}\mbox{Re}\{[F_{1,a}^\Psi]^*F_{1,a}^\Phi\,e^{-im_at}\}+C_{\Phi}\,,
\label{leading1}
\EN
where we added an index $a$ to account for the general case of several particle species \cite{quench}. Notice that these undamped oscillations are a specific dynamical feature of quenches in presence of interaction; as we already remarked, in the free cases $F_n^\Psi\propto\delta_{n,2}$. It is also clear that in the interacting cases these oscillations will be present unless all the one-particle form factors of $\Phi$ and/or $\Psi$ vanish due to an internal symmetry. When this is the case, (\ref{intermediate}) will hold with $n_0>1$, most commonly $n_0=2$ and a suppression by a power $t^{-3/2}$. In all cases $n_0$ can be determined by symmetry considerations, as we will illustrate in the next section through a number of examples. It follows from (\ref{intermediate}) that $C_\Phi$ is the relaxation value within the intermediate time window. We also have
\EQ
\lim_{t\to+\infty}\lim_{\lambda\to 0}\frac{1}{\lambda}\delta\langle\Phi(t)\rangle=\frac{C_\Phi}{\lambda}\,;
\label{asymp}
\EN
for $n_0=1$, $C_\Phi$ has to be intended as the central value of the undamped oscillations of $\delta\langle\Phi(t)\rangle$. 

It is worth stressing that the derivation of the above results does not require that the pre-quench theory (\ref{A0}) is integrable. In particular, equation (\ref{intermediate}) holds in general and determines on symmetry grounds the damping power of the oscillations in the intermediate time window.  
If the pre-quench theory is integrable, on the other hand, the mass spectrum and the form factors are exactly known, so that the above expressions are completely determined analytic predictions for the full time evolution up to $t_\lambda$. From a computational point of view, it is relevant to mention that form factor series are typically rapidly convergent and provide accurate estimates when truncated to the first few terms (see e.g. \cite{review}). 

Some additional conclusions can be obtained in general. Consider the integrated Euclidean connected two-point function
\bea
\label{euclidean}
&& \int d^2x\,\langle\Psi(x,-it)\Phi(0,0)\rangle_c = \int d^2x\,\langle 0|T\,\Psi(x,-it)\Phi(0,0)|0\rangle_c= \\
&& \int_{-\infty}^\infty dx\left[\int_0^\infty dt\,\langle 0|\Psi(x,-it)\Phi(0,0)|0\rangle_c+\int_{-\infty}^0 dt\,\langle 0|\Phi(0,0)\Psi(x,-it)|0\rangle_c\right]\nonumber\,.
\eea
Inserting a resolution of the identity over the complete set of states $|p_1,\ldots,p_n\rangle$ in between the two operators, performing the integrations over $x$ and $t$, and comparing with (\ref{C}) we obtain 
\EQ
C_\Phi=-\lambda\int d^2x\,\langle\Psi(x,-it)\Phi(0,0)\rangle_c\,;
\label{equilibrium}
\EN
since the r.h.s. gives the first order expression for $\langle\Phi\rangle_\lambda-\langle\Phi\rangle_{\lambda=0}$ {\it at equilibrium}, we see that in the limit (\ref{asymp}) $\langle\Phi(t)\rangle$ approaches (or, for $n_0=1$, oscillates around) the equilibrium value $\langle\Phi\rangle_{\lambda}$. The equilibrium one-point function is a constant, and evaluating it in real or imaginary time makes no difference. 

 It is also known \cite{DSC} that the scaling dimension $X_\Phi$ of the operator $\Phi$ can be expressed as
\EQ
X_\Phi=-\frac{1}{2\pi\langle\Phi\rangle}\int d^2x\,\langle\Theta(x,-it)\Phi(0,0)\rangle_c\,,
\label{X}
\EN
where 
\EQ
\Theta(x,t)=2\pi\,g(2-X_\varphi)\,\varphi(x,t)
\label{trace}
\EN
is the trace of the energy-momentum tensor in the theory (\ref{A0}) (here we only consider the case $X_\varphi<2$ of a strictly relevant operator $\varphi$). Equation (\ref{X}) holds as long as the integral converges, and this in turn happens when the operator $\Phi$ does not mix under renormalization (see \cite{DSC} for a detailed discussion). Restricting to this case and recalling (\ref{equilibrium}) we obtain
\EQ
C_\Phi=\frac{\delta g}{g}\,\frac{\langle\Phi\rangle_{\lambda=0}}{2-X_\varphi}\,X_\Phi\,,\hspace{1cm}\mbox{for}\,\,\,\,\Psi=\varphi\,,
\label{scale}
\EN
where $\langle\Phi\rangle_{\lambda=0}=\langle 0|\Phi(0,0)|0\rangle$ is the pre-quench value, and for this specific case we adopted the natural notation $\lambda=\delta g\ll g$.

\section{Examples}
\subsection{Ising chain}
\begin{figure}[t]
\begin{center}
\includegraphics[width=\textwidth]{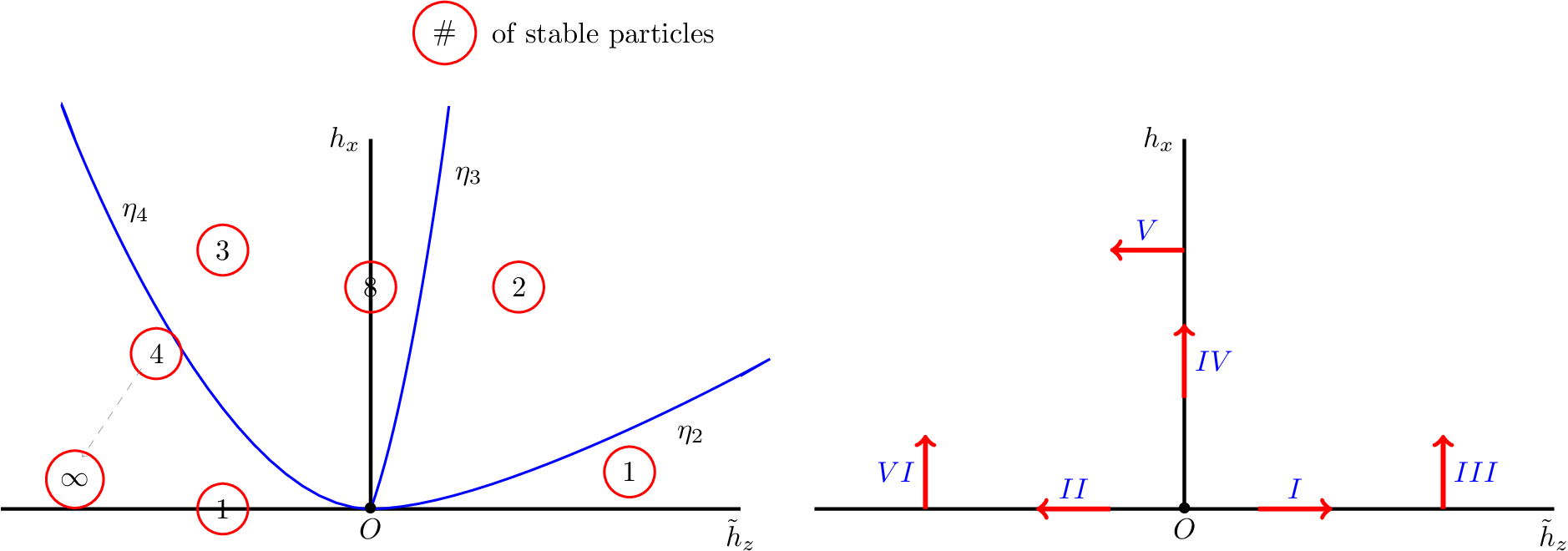}
\caption{\textit{Left.}~Evolution of the mass spectrum of the near-critical Ising chain in coupling space.
The quantum critical point is located at the origin. \textit{Right.}~
Schematic representation of the different quantum quenches starting from the three integrable directions in
the coupling space of the near-critical Ising chain. Arrows can be reversed, this operation corresponding to a
change of the sign of $\lambda$ in (\ref{action})}
\label{Ising_quenches}
\end{center}
\end{figure}

The Ising quantum spin chain with external fields is defined by the Hamiltonian
\EQ
H_\textrm{Ising}=-J\sum_{j=-\infty}^{\infty}[\sigma^x_j\sigma^x_{j+1}+h_z\sigma^z_j+h_x\sigma^x_j]\,,
\label{Ising_chain}
\EN
where $\sigma^{\alpha}_j$ denote Pauli matrices at site $j$, and $h_x$ and $h_z$ are the longitudinal and transverse magnetic field, respectively. For $h_x=0$ and $\tilde{h}_z\equiv h_z-1=0$ the system possesses a quantum critical point around which the scaling limit is described by the Ising field theory (see \cite{review} for a review) with action
\EQ
{\cal A}_\textrm{Ising}={\cal A}_\textrm{QCP}^\textrm{Ising}-\int dt\,dx\,[\tilde{h}_z\sigma^z(x,t)+h_x\sigma^x(x,t)]\,,
\label{Ising}
\EN
where we took the freedom to preserve the notations for the couplings and the operators, despite the fact that the former are now scaling variables, and the latter are no longer matrices and depend on the continuous space coordinate $x$ (which has nothing to do with the longitudinal direction in coupling space) and time $t$. In the language of the classical two-dimensional spin model, $\sigma^z(x,t)$ corresponds to the energy operator $\varepsilon(x,t)$, which is even under the spin reversal symmetry of the critical point and has scaling dimension $1$, while $\sigma^x(x,t)$ corresponds to the spin operator $\sigma(x,t)$, which is odd under spin reversal and has scaling dimension $1/8$. The critical point is a theory of free massless Majorana fermions. The theory (\ref{Ising}) remains free for $h_x=0$, with the fermions acquiring a mass, and a paramagnetic (resp. ferromagnetic) phase corresponding to $\tilde{h}_z$ positive (resp. negative). For $h_x\neq 0$ the theory is interacting, with a mass spectrum which evolves \cite{McW,nonint,ZF} with the value of the dimensionless parameter 
\EQ
\eta=\tilde{h}_z/|h_x|^{8/15}\,.
\EN
Starting from the paramagnetic phase ($\eta=+\infty$), the number of stable particles in the spectrum increases from $n-1$ to $n$ when $\eta$ passes below a threshold value $\eta_n$ (Fig.~\ref{Ising_quenches}). The number of stable particles goes to infinity when approaching the ferromagnetic phase ($\eta\to-\infty$), as a consequence of the fact that the longitudinal field removes the ground state degeneracy of the ferromagnetic phase and confines the kinks into a sequence (dense for $h_x\to 0$) of topologically neutral mesons. The case $\tilde{h}_z=\eta=0$ is special in several respects. It is known to become integrable in the scaling limit and to possess a mass spectrum consisting of eight stable particles with mass ratios whose values appear in the theory of the Lie algebra $E_8$ \cite{Taniguchi}; for this reason this is often referred to as the ``$E_8$'' spectrum. The heavier five of these particles, however, have a mass larger than the lowest decay threshold and are stable only because of integrability; they decay as soon as a small transverse field is switched on \cite{DGM}, leaving only three stable particles. It is indeed known that $\eta=0$ lies in between $\eta_3$ and $\eta_4$. 

%%%%%%%%%%%%%%%%%%%%%%%%%%%%%%%%%%%%%%%%%%%%%%%%%%%%%%%%%%%%%%%
%%%%%%%%%%%%%%%%%%%%%%%%%%%%%%%%%%%%%%%%%%%%%%%%%%%%%%%%%%%%%%%%  

Having recalled the relevant notions of the equilibrium theory, we can move on and consider quantum quenches. We will consider cases in which the pre-quench theory is integrable since, as we explained, this yields additional analytic information; the initial state will be the ground state $|0\rangle$ of the pre-quench theory. There are three integrable directions in the coupling plane ($\tilde{h}_z,h_x$), namely the paramagnetic and ferromagnetic phases at $h_x=0$, and the theory with $\tilde{h}_z=0$. Choosing one of these integrable theories as the term ${\cal A}_0$ in the action (\ref{action}), we can choose the quench term $\lambda\Psi$ to coincide with $\tilde{h}_z\sigma^z$ or $h_x\sigma^x$, so that we can consider the six different quenches indicated in Fig.~\ref{Ising_quenches}. With reference to (\ref{A0}) and (\ref{action}), they are also specified in the first three columns of Table~\ref{Ising_table}. We now separately discuss these quenches.

Quenches I and II are mass quenches ($\lambda=\delta h_z\propto\delta m$) in the free fermionic theory and correspond to exactly solvable cases, so that the expansion in $\lambda$ can in principle be performed at all orders and summed up. This, however, is not our goal, and we are interested instead in checking the consistence of the results at leading order in $\lambda$ with other expansions available for this non-interacting case. $F_n^{\sigma^x}$ is non-zero for $n$ odd in the paramagnetic phase and for $n$ even in the ferromagnetic phase, while $F_n^{\sigma^z}\propto\delta_{n,2}$ in both phases  (see \cite{review}). This fixes the values of $n_0(\Phi)$ given in Table~\ref{Ising_table} and determining the behavior of one-point functions through Eq.~(\ref{intermediate}). In Table~\ref{Ising_table} the cross for $n_0(\sigma^x)$ in quench I corresponds to the fact that in this case the longitudinal magnetization (order parameter) vanishes due to the unbroken spin reversal symmetry. In the other cases oscillations with a single frequency (there is a single mass in the spectrum) are damped by a power $t^{-3/2}$ in the intermediate time window. This agrees with the results obtained\footnote{For $\sigma^z$ the damping power $3/2$ was first found in \cite{BMcD} for arbitrarily large times, but with an initial condition at $t=0$ corresponding to thermal equilibrium.} in \cite{CEF,SE,CEF2}. Indeed, since $F_n^{\sigma^z}\propto\delta_{n,2}$, the one-point functions (\ref{1point}) reduce to the two-particle contribution and are easily calculated using the exactly known two-particle form factors \cite{BKW,review} 
\bea
F_2^{\sigma^z}(\theta_1,\theta_2) &=& -i\alpha m\,\sinh\frac{\theta_1-\theta_2}{2}\,,
\label{f2eps}\\
F_2^{\sigma^x}(\theta_1,\theta_2) &=& i\,\langle 0|\sigma^x|0\rangle\,\tanh\frac{\theta_1-\theta_2}{2}\,,
\label{f2sigma}
\eea
where we used the rapidity parameterization $(E_p,p)=(m\cosh\theta,m\sinh\theta)$ for energy and momentum, $\langle 0|\sigma^x|0\rangle$ is the pre-quench spontaneous magnetization, and $\alpha$ is a non-universal, real and dimensionless constant. The calculations are given in the appendix and the results are shown in Figs.~\ref{quench_sz}. The integral in $dp$ expressing $C_{\sigma^z}$ diverges logarithmically, a feature known from the lattice calculation \cite{CEF2} (see also \cite{RMCKT}), and Fig.~\ref{quench_sz} shows the result for two values of the cutoff; recalling the comment following (\ref{X}) and (\ref{trace}), this divergence is due to the mixing of $\sigma^z$ with the identity operator \cite{DSC}. On the other hand $\sigma^x$ does not mix, and (\ref{scale}) gives 
\EQ
C_{\sigma^x}=\frac{\delta h_z}{8h_z}\langle 0|\sigma^x|0\rangle\,,
\label{Cx}
\EN
a result which also agrees with recent numerical data \cite{Kormos}.

%%%%%%%%%%%%%%%%%%%%%%%%%%%%%%%%%%%%%%%%%%%%%%%%%%%%%%%%%%%%%%%%%%
\begin{table}
\begin{center}
\begin{tabular}{|c||l|c||c|c|c|}
\hline
{\color{blue}{quench}} & \hspace{.7cm}$\varphi$ & $\Psi$ & $n_0(\sigma^x)$ & $n_0(\sigma^z)$ & $\#$ of frequencies \\
\hline\hline
{\color{blue}{I}} & $\sigma^z$ ($\tilde{h}_z>0$) & $\sigma^z$ & $\times $ & $2$ & 1 \\
\hline
{\color{blue}{II}} & $\sigma^z$ ($\tilde{h}_z<0$) & $\sigma^z$ & $2$ & $2$ & 1 \\
\hline
{\color{blue}{III}} & $\sigma^z$ ($\tilde{h}_z>0$) & $\sigma^x$ & $1$ & $\times $ & 1 \\
\hline
{\color{blue}{IV}} & $\sigma^x$ & $\sigma^x$ & $1$ & $1$ & 8 \\
\hline
{\color{blue}{V}} & $\sigma^x$ & $\sigma^z$ & $1$ & $1$ & 8 \\
\hline
{\color{blue}{VI}} & $\sigma^z$ ($\tilde{h}_z<0$) & $\sigma^x$ & $2$ & $2$ & 1\\
\hline
\end{tabular}
\caption{The quantum quenches in the near-critical Ising chain indicated in Fig.~\ref{Ising_quenches}. With reference to (\ref{A0}) and (\ref{action}), $\varphi$ specifies the pre-quench theory, and $\Psi$ the quench operator. $n_0(\Phi)$ is the integer entering (\ref{intermediate}); see the text for the meaning of the crosses. The number of oscillation frequencies (i.e. masses of the pre-quench theory) for small quenches are given in the last column.}
\label{Ising_table}
\end{center}
\end{table}
%%%%%%%%%%%%%%%%%%%%%%%%%%%%%%%%%%%%%%%%%%%%%%%%%%%%%%%%%%%%%%%%%%

Quench III is the first involving a non-zero longitudinal field (and then interaction) and for which field theory yields the only available analytic results. The pre-quench theory is the same as for quench I, and the cross in Table~\ref{Ising_table} indicates that the vanishing of $[F_n^{\sigma^x}]^*F_n^{\sigma^z}$ for all $n$ gives a vanishing first order contribution to the one-point function of $\sigma^z$. On the other hand, since $F_1^{\sigma^x}\neq 0$, $\delta\langle\sigma^x(t)\rangle$ is given for small $\lambda$ by (\ref{intermediate}) with $n_0=1$. The undamped oscillations have been numerically observed in \cite{RMCKT} with the predicted frequency and amplitude. Concerning the constant $C_{\sigma^x}$ which determines the central value,
the $n=1$ contribution to (8) reads $-2h_x|F_1^{\sigma^x}|^2/m^2$. The next
contribution, i.e. that with $n=3$, is obtained using
$F_3^{\sigma^x}(\theta_1,\theta_2,\theta_3)=-iF_1^{\sigma^x}\prod_{i<j}\tanh
\frac{\theta_i-\theta_j}{2}$ \cite{BKW,review}, and turns out to be three order of magnitudes smaller that the $n=1$ term. The fast suppression of
contributions as $n$ increases is typical of form factor series of the type
(8) (see \cite{review}). As a consequence, $C_{\sigma^x}$ essentially coincides with the $n=1$ term, whose absolute value is in turn equal to the amplitude of
the oscillations, as it is seen from (12) for the present case of a single particle species. This non-trivial prediction of the theory is beautifully confirmed
by the numerical data reported in Table~6.1 of \cite{RMCKT}. In particular, for the
three smallest values of longitudinal field $h_x=-0.01,-0.02,-0.03$ the
measured oscillation amplitudes are 0.0368, 0.0732, 0.1087, respectively,
and the measured values of $C_{\sigma^x}$ are 0.037, 0.073, 0.109,
respectively.

Quenches IV and V start from the integrable theory with eight stable particles in the spectrum. Since spin reversal symmetry is broken from the beginning, all the one-particle form factors are non-vanishing (and known \cite{immf,DS,review}) and the one-point functions oscillate at large times according to (\ref{leading1}), with frequencies coinciding with the different masses (the result for the longitudinal magnetization in quench IV is shown in Fig.~\ref{quench_E8}). As a consequence, performing quench IV and looking at one-point functions at sufficiently large times, it should be possible to observe the full ``$E_8$'' spectrum through a Fourier analysis. This is non-trivial in view of the fact that this spectrum is only partially observed at equilibrium, where the signal of the heavier particles is confused with that of the continuum \cite{Coldea}. Equation (\ref{leading1}) is expected to hold, with a sum running over eight masses, also for quench V, despite the fact that at equilibrium a non-zero $\tilde{h}_z$ leads to the decay of the heavier five particles. The point is that within our description of the quench problem the particle states of the pre-quench theory provide the basis on which one computes the post-quench dynamics, and the states of this basis will never be eigenstates of the post-quench Hamiltonian. Hence, the fact that the heavier five particles are unstable within the post-quench theory is immaterial; the circumstance is encoded in the form factors order by order in $\lambda$.

Similar considerations apply to quench VI. Indeed, as we already recalled, at equilibrium a non-zero $h_x$ removes from the spectrum of asymptotic particles the excitations (kinks) of the ferromagnetic phase, which connect states which are no longer degenerate; the new asymptotic excitations are the mesons resulting from kink-antikink confinement. Once again, however, this should not affect the validity of the expansion in the quench parameter, and for $\lambda=h_x$ small enough (\ref{intermediate}) is expected to hold on the kink basis with $n_0=2$ and a single frequency for both $\sigma^x$ and $\sigma^z$. This is not in contradiction with the multi-frequency oscillations found numerically in \cite{KCTC,RMCKT} because those data refer to values of $h_x$ for which the meson masses are already macroscopically separated. The first order results (\ref{1point}), (\ref{intermediate}), on the other hand, are expected to hold for values of $h_x$ small enough to keep the meson spectrum still quite dense, and it would be interesting to test them numerically in this case. Within the expansion in $\lambda$, the oscillation pattern of \cite{KCTC,RMCKT} should arise at higher orders.

%%%%%%%%%%%%%%%%%%%%%%%%%%%%%%%%%%%%%%%%%%%%%%%%%%%%%%%%%%%%%%%%%%%%%%%%%%%%%%%%%%%%%%%%%%%%%%%%%%%%%%%%%%
\begin{figure}[t]
\begin{center}
\includegraphics[width=\textwidth]{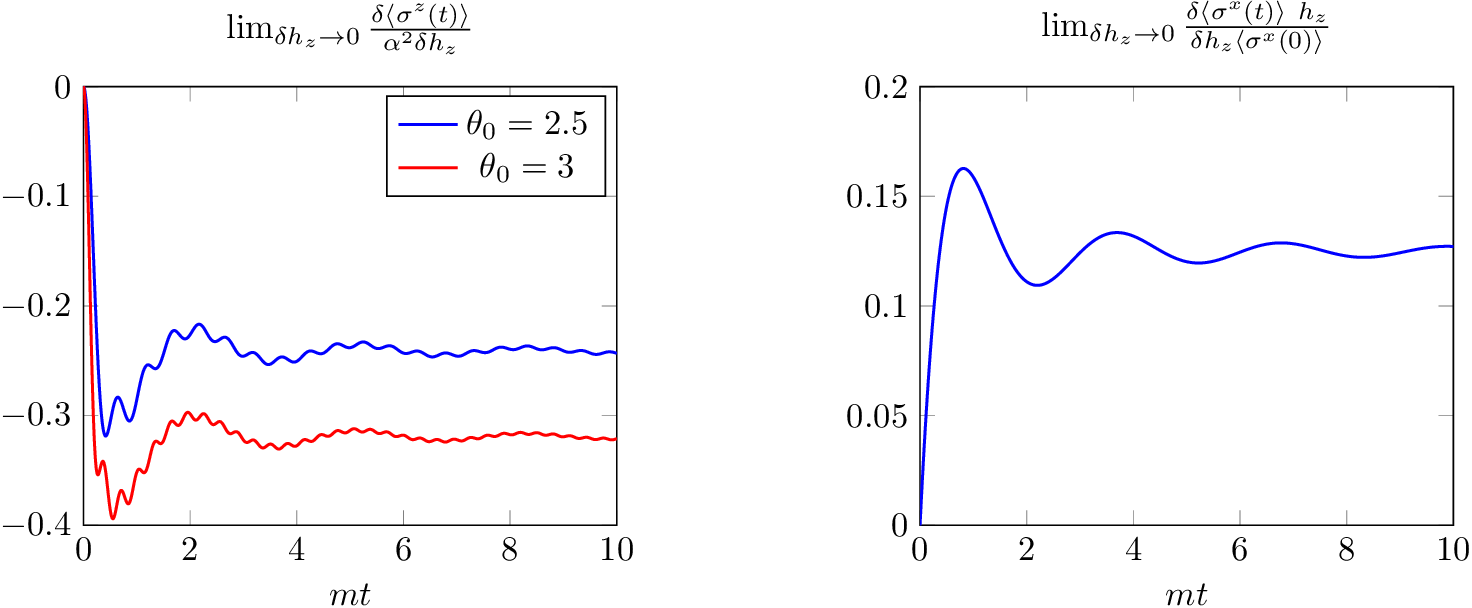}
\caption{\textit{Left.}~$\lim_{\delta h_z\to 0}\delta\langle\sigma^z(t)\rangle/(\alpha^2\,\delta h_z)$
determined by (\ref{1point}) and (\ref{f2eps}) 
for quenches I and II of Fig.~\ref{Ising_quenches} in the near-critical Ising chain. The relaxation value is cutoff dependent (see text) and the two curves correspond to $\int_{-\theta_0}^{\theta_0}d\theta$ evaluated for $\theta_0=2.5$ (upper curve) and $\theta_0=3$. 
\textit{Right.~}$\lim_{\delta h_z\to 0}\delta\langle\sigma^x(t)\rangle\,h_z/(\delta h_z\,
\langle\sigma^x(0)\rangle)$ determined by (\ref{1point}), (\ref{f2eps}), (\ref{f2sigma}) and (\ref{Cx}) for quench II of Fig.~\ref{Ising_quenches} in 
the near-critical Ising chain.}
\label{quench_sz}
\end{center}
\end{figure}
%%%%%%%%%%%%%%%%%%%%%%%%%%%%%%%%%%%%%%%%%%%%%%%%%%%%%%%%%%%%%%%%%%%%%%%%%%%%%%%%%%%%%%%%%%%%%%%%%%%%%%%%%%%%%%%

\subsection{XYZ chain}
The XYZ quantum spin chain is defined by the lattice Hamiltonian
\EQ
H_\textrm{XYZ}=-J\sum_{j=-\infty}^\infty[(1+\gamma)\,\sigma^x_j\sigma^x_{j+1}+(1-\gamma)\,\sigma^y_j\sigma^y_{j+1}+\Delta\,\sigma^z_j\sigma^z_{j+1}]\,,
\label{xyz}
\EN
and for $\gamma=0$ reduces to the XXZ chain. The latter is critical for $|\Delta|<1$ and renormalizes in the continuum limit onto the massless Gaussian model \cite{LP,KB}. It is a consequence of this correspondence that in the scaling limit $\gamma\ll 1$ (\ref{xyz}) is described by the sine-Gordon quantum field theory with action
\EQ
{\cal A}_\textrm{SG}=\frac{1}{16\pi}\int dt\,dx\,[(\partial_t\phi)^2-(\partial_x\phi)^2]-g\int dt\,dx\,\cos\beta\phi\,,
\label{sg}
\EN
where
\EQ
\cos\pi\beta^2=\Delta\,,\hspace{1cm}0<\beta^2<1\,,
\label{delta}
\EN
$g\propto\gamma$, and $\cos\beta\phi\sim\sum_j[\sigma^x_j\sigma^x_{j+1}-\sigma^y_j\sigma^y_{j+1}]$; this operator drives the system away from criticality (i.e. corresponds to $\varphi$ in (\ref{A0})), and we refer to it as the plane anisotropy operator; its scaling dimension is
\EQ
X_{\cos\beta\phi}=2\beta^2\,.
\label{Xcos}
\EN
The sine-Gordon theory is quantum integrable \cite{ZZ} and possesses the soliton $A$ and antisoliton $\bar{A}$ as fundamental excitations interpolating between adjacent minima of the periodic potential; these particles become free fermions at $\beta^2=1/2$. For $0<\beta^2<1/2$ the interaction between $A$ and $\bar{A}$ becomes attractive and gives rise to soliton-antisoliton bound states $B_n$ with masses \cite{ZZ}
\EQ
m_n=2m\,\sin\frac{n\xi}{2}\,,\hspace{1cm}1\leq n<\frac{\pi}{\xi}\,,
\label{mn}
\EN
where $m$ is the soliton mass and 
\EQ
\xi=\frac{\pi\beta^2}{1-\beta^2}\,.
\label{xi}
\EN
The particle $B_1$ is created by the bosonic field $\phi$, and is odd under the $Z_2$ transformation $\phi\to-\phi$ which leaves the action (\ref{sg}) invariant. More generally, it turns out that the particles $B_n$ have parity $(-1)^n$ under this symmetry. The sine-Gordon form factors are also known \cite{KW,Smirnov} (see also \cite{DG}). 

\begin{figure}[t]
\begin{center}
\includegraphics[width=7cm]{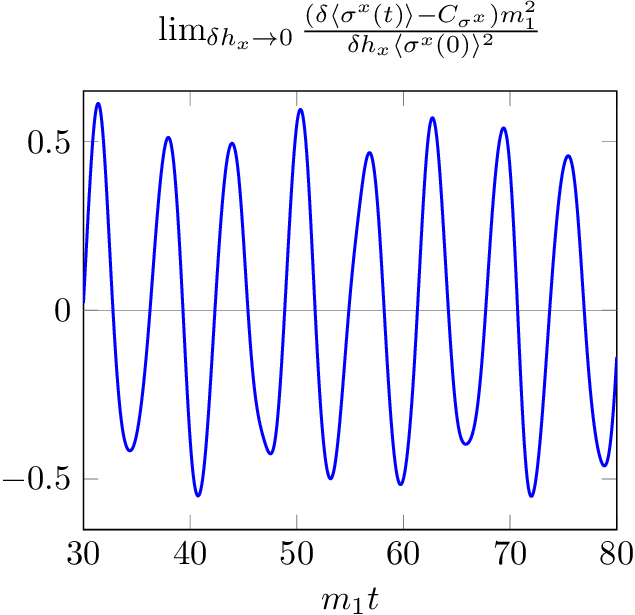}
\caption{$\lim_{\delta h_x\to 0}(\delta\langle\sigma^x(t)\rangle-C_{\sigma^x})\,m_1^2/(\delta h_x\,\langle\sigma^x(0)\rangle^2)$ at large times determined by (\ref{leading1}) and by the known mass spectrum and form factors (see \cite{review}) for quench IV of Fig.~\ref{Ising_quenches} in the near-critical Ising chain. Notice that (\ref{scale}) gives $C_{\sigma^x}=\frac{\delta h_x}{15h_x}\langle\sigma^x(0)\rangle$.}
\label{quench_E8}
\end{center}
\end{figure}

We now consider the quantum quench\footnote{Time evolution in the sine-Gordon model was considered in \cite{GDLP,BES} for initial conditions at $t=0$ expressed in terms of the particle excitations of the $t>0$ theory, without a notion of pre-quench Hamiltonian. This type of problem is not in the class considered in this paper.} of the XYZ chain in which we change the plane anisotropy parameter $\gamma$ by an amount $\delta\gamma\ll\gamma$ at $t=0$. In the scaling limit this corresponds to the theory (\ref{action}) with ${\cal A}_0={\cal A}_\textrm{SG}$, $\lambda\propto\delta\gamma$ and $\Psi=\cos\beta\phi$. For the case in which the system is in the ground state of the pre-quench Hamiltonian for $t<0$, the theory predicts the behavior (\ref{intermediate}) for the one-point functions to first order in $\lambda$, with a value of $n_0$ which in general depends on $\beta$ (i.e. on $\Delta$). In particular, for $\Phi=\cos\beta\phi$ we have $n_0$=2 for $\beta^2\geq 1/3$ and $n_0=1$ for $\beta^2<1/3$. This follows from the fact that $\cos\beta\phi$ is $Z_2$-even and can have non-zero one-particle form factors only on the particles $B_{2k}$. Eqs.~(\ref{mn}) and (\ref{xi}) show that no such particle is present in the spectrum for $\beta^2\geq 1/3$, so that in this range the dominant contribution comes from the $Z_2$-even two-particle state $A\bar{A}+\bar{A}A$. Lowering $\beta^2$ below the value $1/3$, the particle $B_2$ starts to contribute, followed by the other particles $B_{2k}$ as $\beta^2$ becomes smaller and smaller. Hence, for $\beta^2$ small enough one has to use (\ref{leading1}) including all the frequencies (masses) allowed by (\ref{mn}). The relaxation value $C_{\cos\beta\phi}$ can also be determined using (\ref{scale}) and (\ref{Xcos}); in this case (\ref{scale}) applies for $\beta^2<1/2$, where the integral in (\ref{X}) converges (see \cite{DG}).

\section{Quenching form an excited state}
 In this section we consider the case in which the initial state is not the ground state of $H_0$ but an excited state of this Hamiltonian. More precisely, we consider the first excited state, i.e. the single particle state $|q\rangle$. The post-quench state now reads
\begin{equation}
 |\psi_1\rangle\equiv S_\lambda|q\rangle\simeq
 |q\rangle+\lambda\sum_{n=2}^\infty\frac{2\pi}{n!}
 \int_{-\infty}^{\infty}\prod_{i=1}^n\frac{dp_i}{2\pi E_{p_i}}\delta
 \left(\sum_{i=0}^n p_i-q\right)\frac{[F_{1,n}^{\Psi}(q|p_1,\dots, p_n)]^*}{\sum_{i=0}^n
 E_{p_i}-E_{q}-i0}|p_1,\dots, p_n\rangle,
\label{psi1}
\end{equation}
where 
\EQ
F_{1,n}^{\Psi}(q|p_1,\dots, p_n)=\langle q|\Psi(0,0)|p_1,\dots, p_n\rangle\,,
\EN
and $i0$ is the infinitesimal imaginary part already discussed in section~2. We restrict our analysis to the case $q\neq 0$, so that the appearance of the term $n=0$ in (\ref{psi1}) is forbidden by momentum conservation.
The term $n=1$ would contribute an infinity due to the vanishing of the energy denominator, and is subtracted; the first order vacuum contribution does not appear in (\ref{psi0}) for the same reason\footnote{See \cite{nonint} for similar subtractions at equilibrium.} \cite{quench}. As a consequence, the first order variation of a one-point function with respect to its pre-quench value takes the form
\bea
\delta\langle\Phi(t)\rangle_1 &\simeq & \frac{\langle\psi_1|\Phi(x,t)|\psi_1\rangle-\langle q|\Phi(0,0)|q\rangle}{\langle\psi_1|\psi_1\rangle}+D_\Phi\nonumber\\
&=& \frac{\lambda}{\langle q|q\rangle}\sum_{n=2}^\infty\frac{2\pi}{n!}\int_{-\infty}^{\infty}\prod_{j=1}^n\frac{dp_j}{2\pi E_{p_j}}\,\frac{\delta(\sum_{j=1}^np_j-q)}{\sum_{j=1}^nE_{p_j}-E_q-i0}\label{1point1}\\
&\times& 2\mbox{Re}\{[F_{1,n}^\Psi(q|p_1,\ldots,p_n)]^*F_{1,n}^\Phi(q|p_1,\ldots,p_n)\,e^{-i(\sum_{j=1}^nE_{p_j}-E_q)t}\}+D_{\Phi}\,,\nonumber
\eea
where $D_\Phi$ is fixed by the requirement $\delta\langle\Phi(0)\rangle_1=0$, and we took into account that $\langle\psi_1|\psi_1\rangle=\langle q|q\rangle+{O}(\lambda^2)$. Since 
\EQ
\langle q_1|q_2\rangle=2\pi E_{q_1}\delta(q_1-q_2)\,,
\EN
we have $\langle q|q\rangle\propto\delta(0)\propto L$, where $L\to\infty$ is the linear size of the system. It follows that, when considering the integrals in (\ref{1point1}) in the limit of infinite size, we can ignore the contributions of order $L^0$. 

For a generic operator $O$ we have 
\bea
F_{1,n}^O(q|p_1,\ldots,p_n) &=& F_{n+1}^O(\bar{q},p_1,\ldots,p_n)\nonumber\\
&+& \sum_{i=1}^{n}2\pi E_{p_i}\delta(p_i-q)\left(\prod_{j=1}^{i-1}S(q-p_j)\right)F_{n-1}^{O}(p_1,\dots,\hat{p}_i,\dots, p_n)\,,
\label{1,n}
 \eea
where $\hat{p}_i$ means omission of the momentum $p_i$, and $S(q-p_j)$ is the scattering phase that the particle with momentum $q$ produces when bypassing that with momentum $p_j$ on its way towards annihilation of the particle with momentum $p_i$. The terms containing the Dirac delta are disconnected parts, while $\bar{q}$ in the connected part indicates the crossing of the corresponding particle. This crossing operation amounts to an analytic continuation which is more conveniently expressed in the rapidity parameterization in which $q=m\sinh\beta$ and $p_i=m\sinh\theta_i$; then the connected part reads \cite{Smirnov}
\EQ
F_{n+1}^O(\beta+i\pi+i0,\theta_1,\ldots,\theta_n)\,.
\label{crossed}
\EN
The infinitesimal imaginary part $i0$ plays a role in the treatment of the {\it kinematical} poles that the form factors exhibit when two rapidities differ by $i\pi$ \cite{BKW,Smirnov}; the physical role of these poles becomes transparent in the study of phase separation \cite{DV,DSq}.

When (\ref{1,n}) is substituted into (\ref{1point1}), the products of two disconnected parts produce a factor $\langle q|q\rangle$ which cancels that in the denominator. Also, in these terms the scattering phases $S(q-p_j)$ in (\ref{1,n}) appear in the product form $S(q-p_j)S^*(q-p_j)=1$. Hence, the disconnected-disconnected contributions to (\ref{1point1}) reconstruct the result $\delta\langle\Phi(t)\rangle$ obtained quenching from the ground state of $H_0$. Concerning the contributions to (\ref{1point1}) involving two connected parts coming from (\ref{1,n}), the regularization of the kinematical poles should make finite the result of the momentum integrals, with the consequence that at $L=\infty$ the connected-connected contributions should be eliminated by the denominator $\langle q|q\rangle\propto L$. The remaining contributions to (\ref{1point1}) are those involving the product of a connected and a disconnected part. The delta function contained by the latter forces the former on a kinematical pole, and the only way out seems that of subtracting these singular parts. Under this hypothesis, also the connected-disconnected contributions can be ignored in the limit $L\to\infty$.

\section{Conclusion}
In this paper we investigated aspects of the theory of quantum quenches in near-critical one-dimensional systems formulated in \cite{quench}. In particular, we showed that for small quenches below the time scale $t_\lambda$ associated to the quench parameter one-point functions of local operators relax to, or exhibit undamped oscillations around, the post-quench equilibrium expectation value. For quenches of the mass scale we related this (average) relaxation value to the scaling dimension of the operator. While $t_\lambda$ can be made arbitrarily large taking the quench parameter sufficiently small, in presence of interaction near criticality there seems to be no way to follow analytically the time evolution beyond $t_\lambda$.  

An aspect of the theory which is remarkable in the context of non-equilibrium dynamics is its generality. Indeed the formulae it yields apply in the vicinity of any quantum critical point with emergent relativistic invariance, and for a variety of quenches within each near-critical region, a property that we illustrated discussing a number of examples related to the quantum critical point of the Ising chain and to the quantum critical line of the XYZ chain. 

A further noticeable feature of the theory is its ability to capture already at first order in the quench parameter $\lambda$ main qualitative aspects of the dynamics such as the presence of undamped oscillations, their relation with the presence of interaction, as well as the role of internal symmetries. This effectiveness originates from the fact that the theory is built, directly in the continuum limit, on the particle excitations which are the fundamental dynamical degrees of freedom. The action of the scattering operator (\ref{Slambda}) on the initial state directly yields the post-quench state of the system. The analysis then shows that the long time behavior of one-point functions is determined by the lowest energy modes contained in the post-quench state, with the consequent discriminant role played by the presence of single-particle modes. They cannot be present in absence of interaction, and  they couple or not to the different observables depending on internal symmetries. 

The first order in $\lambda$ is expected to be always quantitavely accurate for $\lambda$ small enough. Generically it determines the qualitative behavior also for larger values of $\lambda$, unless the quench drives the theory into a region of parameter space where the particles of the pre-quench theory are unstable, due to decay or confinement. In these cases, as discussed in section~3.1, higher orders in $\lambda$ become relevant. 

We also approached the issue of the dependence on the initial state, considering the case in which the quench does not start from the ground state of the pre-quench Hamiltonian, but from its first excited state. The analysis suggests that the first order variation of one-point functions is essentially the same in the two cases, but is complicated by the non-normalizability of excited states on the infinite line, and involves regularization issues that will deserve further study. Numerical investigations, with built-in finite volume regularization, should be especially useful for comparison.

%\newpage

\vspace{1cm} \noindent \textbf{Acknowledgments.} We thank the authors of \cite{Kormos}, and in particular M. Kormos and G. Takacs, for communicating their data before publication.

\appendix
\section{Appendix}
In this appendix we detail the comparison of our results for the quenches I and II in the Ising chain with results previously available for these free fermionic cases. The only non-vanishing form factor of the quench operator $\sigma^z$ is the two-particle one, so that (\ref{1point}) reduces to the form

\EQ
\delta\langle\Phi(t)\rangle \simeq \frac{\delta{h}_z}{8\pi}\int_{-\infty}^{\infty}\frac{dp}{E^3_{p}}\,
2\mbox{Re}\{[F_2^{\sigma^z}(p,-p)]^*F_2^\Phi(p,-p)\,e^{-2iE_{p}t}\}+C_{\Phi}\,,
\label{1point_ising}
\EN
which produces the plots of Fig.~2 upon substitution of (\ref{f2eps}) and (\ref{f2sigma}). For large times small momenta dominate and for quench II we have in particular
\bea
\langle\sigma^x(t)\rangle &\simeq & \langle 0|\sigma^x|0\rangle+\frac{\delta{h}_z}{8\pi}\int_{-\infty}^{\infty}\frac{dp}{m^3_{p}}\,
2\mbox{Re}\{-\alpha\langle 0|\sigma^x|0\rangle\,\frac{p^2}{m}\,e^{-2im(1+p^2/m^2)t}\}+C_{\sigma^x}\,,\nonumber\\
& = & \langle 0|\sigma^x|0\rangle+\frac{\delta{h}_z\,\alpha \langle 0|\sigma^x|0\rangle}{16\sqrt{\pi}\,m(mt)^{3/2}}\,\cos(2mt-\pi/4)+C_{\sigma^x}\,,
\hspace{.6cm}1/m\ll t\ll 1/\delta{h}_z\,,
\label{1point_sigma}
\eea
where we used $p\simeq m\theta$ for small momentum. This result can be compared with the expression
\EQ
\langle\sigma^x(t)\rangle=\bar{\sigma}[1+\frac{a}{\bar{m}t}-\frac{1-\bar{m}/m}{8\sqrt{\pi}\,(\bar{m}t)^{3/2}}\,\cos(2\bar{m}t-\pi/4)+\cdots]e^{-t/\tau}\,,\hspace{1cm}t\to\infty\,,
\label{dirk}
\EN
obtained in \cite{SE} for a quench from mass $m$ to mass $\bar{m}$; $\bar{\sigma}$ is the equilibrium expectation value of $\sigma^x$ corresponding to post-quench parameters. Writing $\bar{m}=m+\delta m$ and expanding for small $\delta m$, it can be checked that the expressions given in \cite{SE} for $a$ and $1/\tau$ are of order $(\delta m)^2$; hence, recalling also (\ref{equilibrium}), we see that (\ref{dirk}) coincides with the first order result (\ref{1point_sigma}) for 
\EQ
\alpha\,\delta h_z=2\,\delta m\,.
\label{alpha}
\EN

For the transverse magnetization (\ref{f2eps}) and (\ref{1point_ising}) give
\EQ
\delta\langle\sigma^z(t)\rangle \simeq \frac{\alpha^2\,\delta{h}_z}{8\pi}\int_{-\infty}^{\infty}d\theta\,\frac{\sinh^2\theta}{\cosh^2\theta}\,\cos(2mt\cosh\theta)+C_{\sigma^z}\,.
\label{1p_eps}
\EN
This result can be compared with the expression
\bea
\langle\sigma^z(t)\rangle &=& -\frac{1}{L}\sum_n[\cos\beta_n\cos\Delta_n+\sin\beta_n\sin\Delta_n\cos(2E_nt)]\,,
\label{CEFlattice}\\
\beta_n &=& \arctan(\sinh\theta_n)\,,\\
\Delta_n &=& \beta_n-\arctan(\bar{m}/m\,\sinh\theta_n)\,,\\
E_n &=& \bar{m}\cosh\theta_n\,,\\
p_n &=& \bar{m}\sinh\theta_n=2\pi n/L \label{pn}
\eea
obtained from the scaling limit \cite{RMCKT} of the lattice result \cite{CEF2} in the paramagnetic phase; $L\to\infty$ is the system size. (\ref{CEFlattice}) can be rewritten as
\bea
\langle\sigma^z(t)\rangle &=& -\frac{1}{L}\sum_n[\frac{\cos\Delta_n}{\cosh\theta_n}+\sin\Delta_n\frac{\sinh\theta_n}{\cosh\theta_n}\cos(2E_nt)]\nonumber\\
&\simeq &-\frac{1}{L}\sum_n[\frac{1}{\cosh\theta_n}-\frac{\delta m}{m}\frac{\sinh^2\theta_n}{\cosh^3\theta_n}\cos(2E_nt)]+O((\delta m)^2)\nonumber \\
&\simeq & \frac{m}{2\pi}\int_{-\infty}^\infty d\theta\,\left\{\frac{\delta m}{m}\left[\frac{\sinh^2\theta}{\cosh^2\theta}\cos(2mt\cosh\theta)-1\right]-1\right\}+O((\delta m)^2)\,,
\eea
where we used (\ref{pn}) to pass to the integral form. Subtracting $\langle\sigma^z(0)\rangle$ and recalling (\ref{alpha}) we recover (\ref{1p_eps}) with a value $\alpha=2$ of the non-universal operator normalization.

\end{document}